\documentclass[aps,prl,twocolumn,superscriptaddress]{revtex4-2}

\usepackage{graphicx}
\usepackage{xcolor}
\usepackage[acronym]{glossaries}
\usepackage{gensymb}

\usepackage{mhchem}

\newcommand{\figwidth}{1.0\columnwidth}

\begin{document}
\title{Autonomous microARPES}

\author{Steinn Ymir Agustsson}
\affiliation{Department of Physics and Astronomy, Aarhus University, 8000 Aarhus C, Denmark}

\author{Alfred J. H. Jones}
\affiliation{Department of Physics and Astronomy, Aarhus University, 8000 Aarhus C, Denmark}

\author{Davide Curcio}
\affiliation{Department of Physics and Astronomy, Aarhus University, 8000 Aarhus C, Denmark}

\author{Søren Ulstrup}
\affiliation{Department of Physics and Astronomy, Aarhus University, 8000 Aarhus C, Denmark}

\author{Jill Miwa}
\affiliation{Department of Physics and Astronomy, Aarhus University, 8000 Aarhus C, Denmark}

\author{Davide Mottin}
\affiliation{Department of Computer Science, Aarhus University, 8000 Aarhus C, Denmark}

\author{Panagiotis Karras}
\affiliation{Department of Computer Science, Aarhus University, 8000 Aarhus C, Denmark}

\author{Philip Hofmann}
\email{philip@phys.au.dk}
\affiliation{Department of Physics and Astronomy, Aarhus University, 8000 Aarhus C, Denmark}

\begin{abstract}
Angle-resolved photoemission spectroscopy (ARPES) is a technique used to map the occupied electronic structure of solids. Recent progress  in X-ray focusing optics has led to the development of ARPES into a microscopic tool, permitting the electronic structure to be spatially mapped across the surface of a sample. This comes at the expense of a time-consuming scanning process to cover not only a three-dimensional energy-momentum ($E, k_z, k_y$) space but also the two-dimensional surface area. Here, we implement a protocol to autonomously search both $\mathbf{k}$- and real space in order to find positions of particular interest, either because of their high photoemission intensity or because of sharp spectral features. The search is based on the use of Gaussian process regression and can easily be expanded to include additional parameters or optimisation criteria. This autonomous experimental control is implemented on the SGM4 micro-focus beamline of the synchrotron radiation source ASTRID2. 
\end{abstract}
\maketitle

\section*{Introduction}

Advanced computational tools and artificial intelligence (AI) are having a major impact on the entire scientific process. AI is being integrated in the generation of hypotheses, performing experiments and discovering patterns in the resulting data  \cite{Krenn:2022aa,Wang:2023aa}. Finding such trends in large, simulated data sets, for example for the discovery of new materials, has been particularly successful \cite{Liu:2017ab}, partly because such data sets are ``clean'' in the sense of being noise-free and because they are easily adaptable to the FAIR data principles (findable, accessible, interoperable, reusable) \cite{Wilkinson:2016aa} that are essential for applying AI tools.

The use of AI for the autonomous control of experiments, on the other hand, promises huge gains in efficiency but it is less developed because the AI tools need to be closely integrated with a complex experimental control environment. Nevertheless, considerable progress has been made. For example, AI can help to control complex experimental setups that only operate in a small volume of a multidimensional parameter space \cite{Jalas:2021aa}.  AI can also guide a scientist in the selection of experimental parameters in order to search for ``interesting'' features. In an experiment such as neutron scattering, collecting a single data point can take a considerable amount of time and AI can help to decide the set of parameters for which the next data point should be taken, avoiding redundant measurements and maximising the information gain in the presence of the data that has already been collected. This is particularly relevant if a high-dimensional parameter space is to be explored.
Indeed, the dramatic increase of possible parameter combinations in a multi-dimensional parameter space can potentially slow down any experiment to a point where AI-controlled parameter selection becomes essential. 

Theoretical concepts for suggesting new parameters have been tested by simulations based on existing large experimental data sets.
In this approach, a ``measurement'' for a set of parameters is simulated by selecting the previously measured experimental data for these parameters and  by then using an AI method to find the set of parameters for the next ``measurement''.
The quality of the data ``measured'' in this way can then be tested against the already available complete experimental data set and, given that experiments probing several parameters are often performed on an (inefficient) regular grid of data points, large gains can be obtained in these simulations \cite{Kanazawa:2019aa,Ziatdinov:2020aa,Melton:2020aa}. Tests of this kind are limited by the data sets they are based on, which need to be measured sufficiently well across the entire parameter space, precluding high-dimensional parameter spaces for which this would simply take too much time. The autonomous control of actual experiments has also been demonstrated, mostly in two-dimensional parameter spaces \cite{Scarborough:2017aa,Noack:2019aa,Noack:2020ab,Noack:2020aa,Parente:2023aa}.
Such implementations need to include a number of considerations beyond maximising the information gain from a next measurement, \emph{e.g.}, the time it takes to change parameters like sample position or temperature, or the radiation damage to sensitive samples exposed to a highly focused beam of X-rays \cite{Scarborough:2017aa}.

When using an algorithm to autonomously suggest new parameters for an experiment, it is not \emph{a priori} clear what criteria should be applied for choosing these parameters. Most commonly, the aim is to optimise the information gain in relation to the data points already taken. This can be achieved by generating new trial parameters through techniques such as Gaussian Process Regression (GPR) \cite{Noack:2021aa}.
For an experiment in which some type of intensity is measured, \emph{e.g.}, the intensity of photoelectrons or scattered neutrons, this results in parameter suggestions that avoid redundant measurements and are more likely to fall into regions of high intensity than into regions with just a weak background \cite{Parente:2023aa}. However, it is not evident that merely looking for the most intense features is the best way to maximise information gain, as one might miss, \emph{e.g.},  important fine structure in low intensity regions.

\begin{figure}[htbp]
  \includegraphics[width=\figwidth]{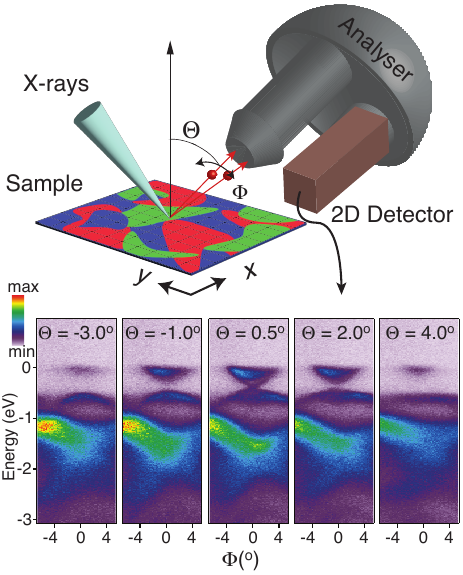}
  \caption{Sketch of a micro-focus ARPES experiment. A narrowly focused X-ray beam is directed towards the surface of a sample, leading to the photoemission of electrons. The X-ray beam can be moved across the sample to inspect different regions, \emph{e.g.} following the grid drawn on the sample surface. In each position, \emph{i.e.}, in each square on the grid, an ARPES spectrum (photoemission intensity as a function of photoelectron kinetic energy and emission angle $\Phi$) can be taken. In addition, the emission angle $\Theta$ can be varied by applying an electrostatic potential in the analyser lens system. The spectra are characteristic of the underlying electronic structure of the sample and will typically be different for sample regions with different characteristics (here indicated by the colour on the sample surface: red, blue and green). The bottom row shows a series of spectra acquired at a fixed position of the sample for different emission angles $\Theta$.   }
  \label{fig:setup_sketch}
\end{figure}

In this paper, we address the challenge of performing autonomous experiments in a multi-dimensional parameter space (measuring the photoemission intensity of a sample in a five-dimensional parameter space and optimising it in a three-dimensional parameter space) and we simultaneously explore the optimisation using more complex criteria than just the intensity. We realise this by applying autonomous experimental control to an ARPES experiment at the micro-focus SGM4 beamline \cite{Volckaert:2023aa} of the ASTRID2 synchrotron radiation facility  in Aarhus, Denmark \cite{Bianchi:2023te}, based on GPR as implemented in the gpCAM package \cite{Noack:2021aa,Noack_gpCAM_2022}. 

A brief sketch of the ARPES experiment is shown in Fig. \ref{fig:setup_sketch}. A narrowly focused X-ray beam with a diameter of a few $\mu$m hits the surface of a sample, causing the emission of photoelectrons. Some of the photoelectrons reach a hemispherical electron analyser, are sorted with respect to kinetic energy and the emission angle $\Phi$ (along a slit in the electrostatic lens system) and captured by a two-dimensional detector. Typical two-dimensional spectra are shown on the bottom of Fig. \ref{fig:setup_sketch}. They give the photoemission intensity $I$ as a function of energy $E$ and emission angle $\Phi$ and  essentially represent an image of a cut through the sample's spectral function \cite{Damascelli:2003aa,Hofmann:2009ab}. The spectral function is of fundamental importance for the properties of solids, encoding features like the velocity and effective mass of the electrons, fingerprints of superconductivity, and many others \cite{Sobota:2021aa}. In order to map the spectral function in the full two-dimensional momentum space, the emission angle perpendicular to the slit, $\Theta$, can be varied by applying an electrostatic potential inside the analyser lens tube. This leads to the series of photoemission spectra in Fig. \ref{fig:setup_sketch} from which the photoemission intensity $I(E,\Phi,\Theta)$ can be inferred. The spectral function can be obtained from this by mapping the emission angles  $(\Phi,\Theta)$ to the crystal momentum $\mathbf{k} = (k_x, k_y)$.

ARPES can be turned into a microscopic technique when tightly focusing the X-ray beam such that the measured spectrum is representative of the small area hit by the beam \cite{Rotenberg:2014aa,Ulstrup:2019aa,Ulstrup:2020aa}. For an inhomogeneous sample consisting of different structures or compositions, as indicated by the colours in the sketch of the sample in Fig. \ref{fig:setup_sketch}, each of these areas will give rise to a characteristic spectrum. Depending on the spatial resolution, this approach is called microARPES or nanoARPES. In such an experiment, the task is thus to determine $I(E,\Phi,\Theta)$ for every point $(x,y)$ on the sample surface or at least for the most ``interesting'' locations on the surface. The parameter space is thus five-dimensional $(E,\Phi,\Theta,x,y)$ but the scanned parameter space it is often only three-dimensional $(\Theta,x,y)$ because the outcome of any measurement is already an image $I(E,\Phi)$ (it might, however, be necessary to explore $\Phi$ over a wider range or to change the range of $E$). Other parameters that are commonly varied in an ARPES experiment are the photon energy \cite{Hufner:2003aa}, the sample temperature \cite{Hofmann:2009ab}, the time delay in pump-probe time-resolved ARPES \cite{Lloyd-Hughes:2021aa}, as well as gating voltages   \cite{Joucken:2019aa,Nguyen:2019aa,Muzzio:2020aa} and currents \cite{Curcio:2020aa,Curcio:2023aa} when studying electronic devices \emph{in operando} \cite{Hofmann:2021tj}. It is thus clear that ARPES is an ideal test case for autonomous experiments in multi-dimensional parameter spaces so large that they cannot be explored without the aid of AI techniques. 

For microARPES, the most commonly employed method to explore the entire surface area of interest is to define a rectangular grid on the surface, as indicated in Fig. \ref{fig:setup_sketch}, and to perform a raster-scan on this grid, collecting an ARPES spectrum in every position of the sample \cite{Nguyen:2019aa,Muzzio:2020aa,Curcio:2020aa}. Once an interesting area has been identified, raster-scans with smaller steps in $x,y$ are performed on this area. While this approach has the advantage of completeness, it also has several drawbacks. Most importantly, performing the raster-scan on a fine grid can be extremely time-consuming and an overview of the entire measured region is only obtained upon completion of the scan. Also, current implementations of the approach only explore $I(E,\Phi,x,y)$ without a variation of $\Theta$. This reduces the dimensionality of the parameter space and speeds up the process but can lead to important details being missed, as we shall illustrate below. A major disadvantage of raster scans is that the time this takes becomes prohibitively long in high-dimensional parameter spaces.  

An alternative approach is to explore the parameter space using randomly selected points. This has the advantage of being able to roughly cover the entire region of interest in a shorter time. The point density (resolution) increases with measurement time such that a coarse overview is already available at an early stage but the measurement can be allowed to proceed until sufficient detail is available. To the best of our knowledge, such random scans are not currently implemented for microARPES measurements, presumably because the irregular point distribution makes it harder to quickly visualise the results. Also, the time needed for moving the sample with respect to the X-ray beam between randomly selected points can be considerable (in our case, this typically takes 10-1000~ms depending on distance between the points). Other drawbacks of random sampling are redundancy and that one could potentially make use of the information contained in already measured points in order to select the next ones. 

GPR, by contrast, is a sophisticated formalism for  interpolating un-sampled data points. This can be used to suggested  new points to be measured, based on the existing data, and it thus forms a basis for performing autonomous experiments. However, due to the need to smoothly integrate GPR guidance into the typically very complex experimental control software, there are so far only few cases in which this has been used in a practical context \cite{Kanazawa:2019aa,Ziatdinov:2020aa,Melton:2020aa,Noack:2019aa,Noack:2020ab,Noack:2020aa,Parente:2023aa}. GPR-based experimental guidance as implemented in the gpCAM package \cite{Noack:2021aa,Noack_gpCAM_2022} is the approach we use here for the autonomous control of a microARPES experiment.

\section*{Experimental Details}

ARPES measurements were performed on the SGM4 micro-focus beamline of ASTRID2 \cite{Volckaert:2023aa,Bianchi:2023te}. As a sample, a Nb-doped Bi$_2$Se$_3$ crystal was used. Crystals of this type show a large variation of co-existing chemical compositions and structural modifications \cite{Kevy:2021aa}. A clean surface was prepared by cleaving in the experimental chamber's load lock at a pressure of $1\times 10^{-8}$~mbar. As we shall see below, the resulting sample surface is highly inhomogeneous. It contains crystalline domains of high quality but variations happen on a sub-$100$~$\mu$m length scale. ARPES measurements were performed at room temperature with a photon energy of 18~eV using a Specs Phoibos 150 SAL analyser. The sample position could be controlled by a 6-axis piezoelectric stick-slip manipulator, capable of repeatable positioning up to 100~nm, with the movement speed of the sample limited to 0.3~mm/s to ensure stability during the measurement. The spatial resolution of the experiment is determined by X-ray focusing and was better than $7$~$\mu$m.

GPR-based autonomous experiments were first performed in a two-dimensional parameter space by varying only the  $(x,y)$ position of the sample in relation to the X-ray beam and subsequently in a three-dimensional parameter space by also varying $\Theta$ via the deflector inside the electrostatic lens system of the analyser. For the two-dimensional parameter space, it was still possible to explore the sample surface by a conventional raster scan. This was done in order to provide a ``ground truth'' for testing and illustration purposes. For the three-dimensional parameter space, such maps were also measured but only for selected values of $\Theta$. A grid spacing of  5~$\mu$m was used to capture fine details.

\section*{Autonomous Data Acquisition}

\begin{figure}[htbp]
  \includegraphics[width=\figwidth]{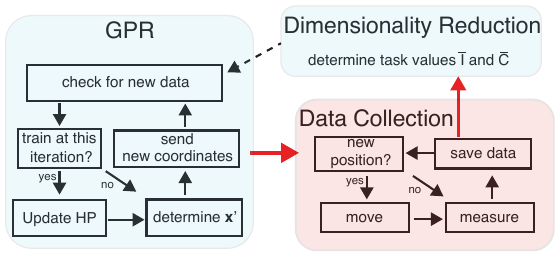}
  \caption{
    Diagram illustrating the GPR-guided data acquisition. Each coloured box represents a process running asynchronously. The arrows represent the flow of data between the processes. HP stands for hyperparameters and $\mathbf{x}^{\prime} = (x^{\prime},y^{\prime},\Theta^{\prime})$ is the new set of parameters to be measured. The red box contains the beamline data acquisition software and the blue boxes are the added GRP implementation. The red arrows indicate the communication between the beamline LabView control software and the Python scripts running the autonomous data acquisition.
  }
  \label{fig:diagram}
\end{figure}

The implementation of autonomous data acquisition is illustrated in Fig. \ref{fig:diagram}. It extends over a three-dimensional parameter space, the position on the sample $(x,y)$ and the emission angle $\Theta$. For convenience, we summarise these in the parameter vector $\mathbf{x} = (x,y,\Theta)$. It is straight-forward to extend  $\mathbf{x}$ such as to include additional parameters.  The autonomous data acquisition consists of three independent processes running asynchronously: (1) the actual measurement at a given  position $\mathbf{x}$; (2) the dimensionality reduction used to extract the quantities that are to be optimised in the GPR-guided search, \emph{e.g.}, the mean photoemission intensity; and (3) the process implementing the GPR-based suggestion of a new parameter set $\mathbf{x}^{\prime}$ for collecting the next data point. Keeping the three processes independent ensures that they can keep running even when one process is locked in a waiting status, such as when receiving or transmitting data. Also, from a practical point of view, the separation of processes implies a minimal need to interfere with the existing measurement software. In the following, each process and its interaction with the others is described in detail.

\subsection*{Data Collection}

The data collection runs off of the existing LabView control software of the microARPES end-station. In order to start the autonomous data acquisition, one or several points in parameter space are chosen, and measurements are performed for these points. Such points can for example be placed in the center and at the boundaries of  the parameter space to be explored. The spectra $I(E, \Phi)$ collected at these points serve to initialise the GPR and result in the first suggested new point in parameter space that is then measured. After each measurement, the data is stored, handled by the dimensionality reduction, and becomes available to the GPR. Once a measurement is completed, the data collection script checks if a new parameter set $\mathbf{x}^{\prime}$ has become available from the GPR script. If yes, the sample is moved to this new position and the next measurement is initiated. If not, another spectrum is collected in the same position, improving the statistics of the data while waiting for a new position. In the remainder of the paper, the number of measured points $N$ refers to measurements at unique parameter set $\mathbf{x}$. Re-measurements at a given parameter set due to the process waiting for a new $\mathbf{x}^{\prime}$ are not counted. A new parameter set provided by the GPR is  approximated to the closest point on the pre-defined grid with a point spacing of 5~$\mu$m for the spatial dimensions and 1\degree~ for $\Theta$. Several software limits are in place to prevent the parameters from moving outside a pre-defined parameter range.  The communication with the GPR takes place via a transmission control protocol (TCP).

\subsection*{Dimensionality Reduction}

The GPR is tasked with suggesting the next ``most promising'' parameter set $\mathbf{x}^{\prime}$ based on the previously collected information. The dimensionality reduction provides the quantities that can be used to evaluate what is ``most promising''. These quantities are referred to as \emph{tasks} \cite{Bonilla:2007aa}. In previous experimental implementations of GPR-controlled experiments, the only task has been the measured intensity or count rate collected for a certain set of experimental parameters, arguing that locations of high intensity / count rate provide most information. In ARPES, the corresponding task is the photoemission intensity integrated over an entire $I(E,\Phi)$ spectrum. Indeed, since a very low photoemission intensity is observed when the X-ray beam does not hit the actual sample, this intensity is often used to locate the sample position as a first step in any experiment. The integration of $I(E,\Phi)$ can also be confined to a certain region of interest (ROI), to exclude experimental artefacts at the edges of the spectra, to choose, for example, just a small energy range around a particular core level energy if only materials containing the corresponding element are of interest, or to restrict the energy window around the Fermi energy $E_F$ for finding the metallic parts of the sample. 

However, due to the two-dimensional detector, ARPES does not only give one single value for the intensity for a given set of parameters, but rather a spectrum $I(E,\Phi)$ and it is thus possible to define additional tasks that take advantage of the information contained in these spectra. Indeed, $I(E,\Phi)$ has the character of an image and a multitude of quantities for measuring the information content of images is readily at hand, for example the Shannon entropy or the edge density. Here we restrict ourselves to the tasks that are already well-established in the ARPES community: The integrated intensity inside a ROI, the magnitude of the integrated second derivative, and the magnitude of the integrated curvature, as defined in Ref. \cite{Zhang:2011aa}. The second derivative and the curvature are often used to make weak structure on a high background more clearly visible. In the present context, they can also serve as a measure of how rich a spectrum is in terms of sharp structures. 

\begin{figure}[htbp]
  \includegraphics[width=\figwidth]{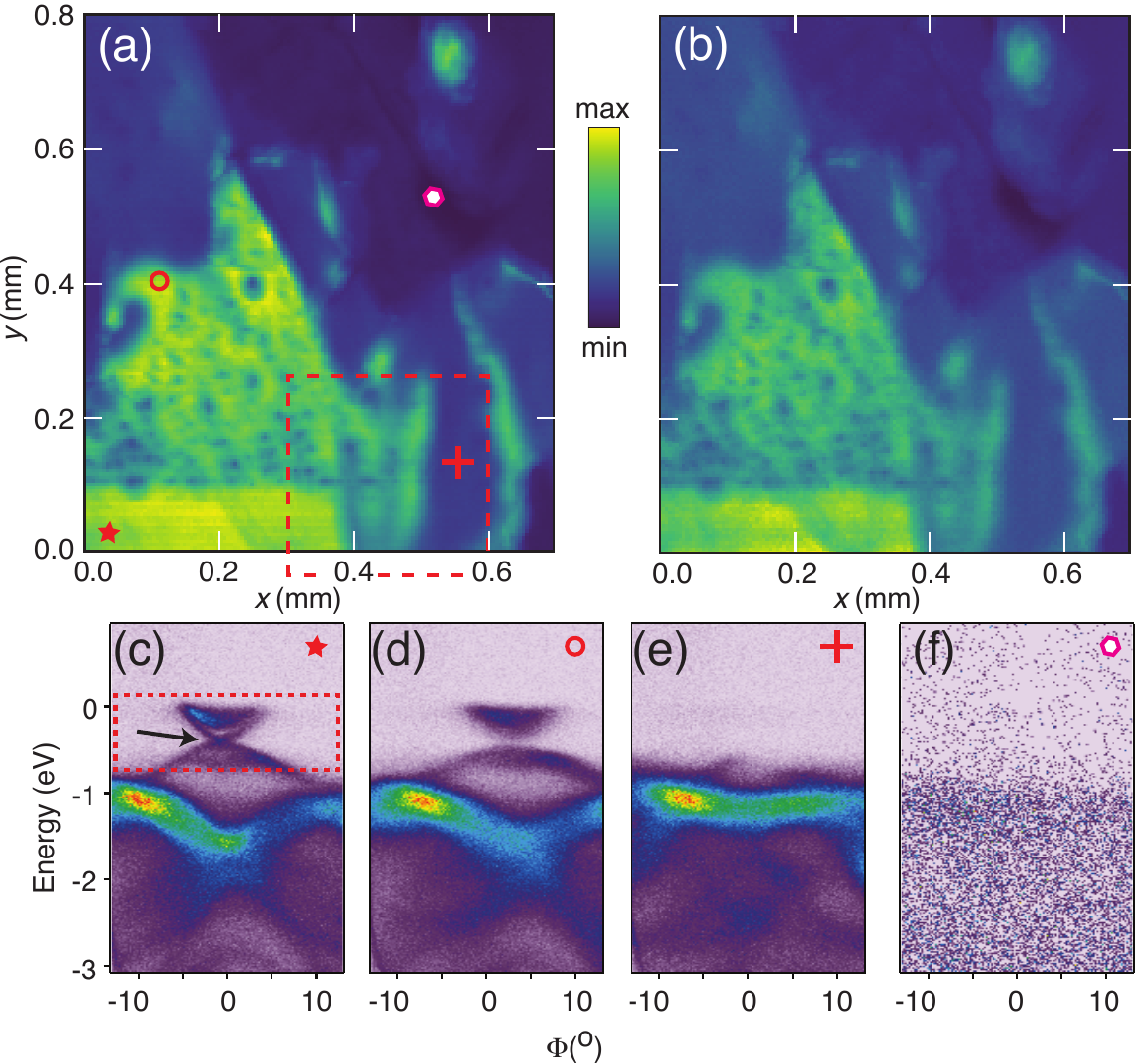}
  \caption{Systematic sample exploration as a benchmark for the GPR-controlled experiment. (a) Mean photoemission intensity $\overline{I}$ across the sample surface. The markers show the locations for the example spectra in panels (c) - (f). The dashed rectangle is the area used for the three-dimensional GPR-controlled data collection. (b) Mean curvature $\overline{C}$ across the sample. (c) - (f) Example spectra taken at the locations with the corresponding markers in panel (a). The dashed rectangle in (c) marks the ROI used for the evaluation of $\overline{I}$ and $\overline{C}$. 
  }
  \label{fig:ground_truth}
\end{figure}

After defining a suitable ROI, the mean intensity $\overline{I} $ of a spectrum inside the ROI is defined as the sum of the intensities of each pixel $I(E_i,\Phi_j)$, normalised by the number of pixels $N_p$ in the ROI:
\begin{align}
  \overline{I} = \frac{1}{N_p} \sum_{E_i,\Phi_j} I(E_i,\Phi_j).
\end{align}
The mean intensity for our sample is shown in Fig. \ref{fig:ground_truth}(a), for a ROI outlined in Fig. \ref{fig:ground_truth}(c). 

The second derivative and the curvature of the photoemission images $I(E,\Phi)$ lead to the definition of two additional tasks. 
The mean sharpness $\overline{S}$ is defined as the normalised sum of the second derivatives of the spectra along the energy and angle directions
\begin{align}
  \overline{S} = \frac{1}{N_p} \sum_{E_i,\Phi_j} \left( \partial^2_{EE} + \partial^2_{\Phi \Phi}  \right),
\end{align}
where a  we introduce the following shorthand for partial derivatives taken at a given point:
\begin{equation}
\partial^2_{EE} =\left. \frac{\partial^2 I(E,\Phi)}{\partial E^2} \right|_{E_i,\Phi_j},
\end{equation}
and correspondingly for other second and first derivatives. 

The mean curvature $\overline{C}$ is defined following Ref. \cite{Zhang:2011aa} as 
\begin{widetext}
\begin{align}
  \overline{C} = \frac{1}{N} \sum_{E_i,\Phi_j} \frac{\left[ 1+c_E (\partial_E)^2\right] c_{\Phi} \partial^2_{\Phi \Phi} - 2 c_{E} c_{\Phi} \partial_E \partial_{\Phi} \partial^2_{E \Phi} + \left[ 1+c_{\Phi} (\partial_{\Phi})^2\right] c_{E} \partial^2_{EE}}{\left[ 1+c_E(\partial_E)^2 + c_{\Phi} (\partial_{\Phi})^2 \right]^{3/2} },
\end{align}
\end{widetext}
where $c_E$ and $c_{\Phi}$ are positive constants (chosen to be 0.001). To avoid noise-generated artefacts, the $I(E,\Phi)$  images are convoluted with a box kernel before calculating $\overline{C}$.

Fig. \ref{fig:ground_truth}(b) shows $\overline{C}$ corresponding to $\overline{I}$ in panel (a) of the same figure. $\overline{C}$ is extremely similar to $\overline{S}$ and therefore the latter is not shown here (in fact,  in the limit of  $c_E=c_{\Phi}=0$, $\overline{C}$ becomes equal to $\overline{S}$). When using $\overline{S}$ or $\overline{C}$ as tasks, one might decide to choose a ROI that does not include the Fermi energy $E_F$ to avoid artefacts due to the sharp cut-off in the Fermi-Dirac distribution. Due to the relatively high sample temperature here (room temperature), this was not found to be necessary. In the remainder of the paper, we use $\overline{I}$ and $\overline{C}$ as tasks. 

\subsection*{Gaussian Process Regression}

The GPR script is the core of the autonomous experiment, responsible for evaluating the hitherto collected data. Gaussian processes (GPs) are non-parametric models and, as such, they rely on special types of similarity functions, called kernels. Once a specific kernel function is chosen, its detailed characterisation is given by a set of so-called hyperparameters. The GPR script cycles between training the GP, \emph{i.e.}, updating the hyperparameters of its kernel, and evaluating the next position to measure. The GPR is implemented as a Python script  based on the GPR-driven data acquisition described in Ref. \cite{Noack:2021aa} and uses the Python library gpCAM \cite{Noack_gpCAM_2022}. 

From a computational point of view, the training of the GP is the most expensive step. The time needed for training depends on the number of available data points and the number of hyperparameters, \emph{i.e.}, the number of parameters that are needed to completely describe the kernel of the GP. We use the default Mat\'{e}rn kernel in gpCAM which is a stationary and anisotropic kernel. This type of kernel is well-suited for the present purpose because it is able to describe the abrupt changes between different sample regions. The number of hyperparameters depends on the number of tasks to be optimised and the dimensionality of the input space. In our case of 2 tasks in a 3(2)-dimensional parameter space, the number of hyperparameters is 5(4). In our experiments, the number of data points per autonomous run never exceeds 1000, leading to a training time of less than a minute on a conventional personal computer. A re-training of the GP is considered once new data becomes available from the dimensionality reduction process. In order to save time, the GP is only retrained at a pre-defined number of available data points: 10, 20, 50, 100, 200, 500$\dots$, excluding the initialisation points.

Whether or not the GP is re-trained, the detection of a new data set triggers the evaluation of a new parameter suggestion $\mathbf{x}^{\prime}$ via the calculation of the acquisition function $f_a(\mathbf{x})$ divided by the cost function $f_c(\mathbf{x}, \mathbf{x}_{c}, \{\mathbf{p}_i\} )$  according to 
\begin{align}
 \mathbf{x}^{\prime} = \arg \max_{\mathbf{x}} \frac{f_{a}(\mathbf{x})}{f_c(\mathbf{x}, \mathbf{x}_{c}, \{\mathbf{p}_i\} )},
  \label{eq:next_position}
\end{align}
where $\mathbf{x}_{c}$ is the parameter vector the experiment currently is set to and $ \{\mathbf{p}_i\} $ and are all the previously measured parameter vectors. 
The suggested parameter set $\mathbf{x}^{\prime}$ is then made accessible to the data acquisition. The acquisition function $f_a$ is based on the result of the GPR and it is updated in the GPR script. The cost function $f_c$ accounts for the time needed to set new parameters as proposed by the acquisition function and imposes a penalty if this time is long. It also prevents measurements at positions for which data has already been collected. Both functions are discussed in more detail below. 

\subsubsection*{Acquisition Function}

The acquisition function is designed to predict the best position for a next measurement. It is a result of applying the GPR to the existing data. The acquisition function used here is a weighted sum of the posterior mean and variance of the GPR, for each of the tasks used to train the GP ($\overline{I}$ and $\overline{C}$ in our case)
\begin{align}
  f_{a}(\mathbf{x}) = \sum_{t}^{} \beta_t \left( m_t(\mathbf{x}) + \alpha~\sigma_t^2(\mathbf{x}) \right),
  \label{eq:our_acqisition_function}
\end{align}
where the sum is carried out over the two tasks, $m_t(\mathbf{x})$ and $\sigma_t^2(\mathbf{x})$ are the mean and the variance of the GP posterior and $\alpha$ is a parameter which controls the relative influence of mean and variance. For a high $\alpha$, the acquisition tends to favour unexplored regions of parameter space, in which the variance is high. For a small $\alpha$, on the other hand, regions of an already established high $m_t(\mathbf{x}) $ are preferred. One therefore speaks of $\alpha$ controlling the balance between \emph{exploration} and \emph{exploitation}. $\beta_t$ are the weights of each task. For two tasks, the acquisition function requires two parameters that need to be set by the user in order to control the autonomous search process ($\alpha$ and the ratio $\beta_1 / \beta_2$). These parameters are only set once and not updated later. We choose to give the same weight to both tasks throughout this paper.

\subsubsection*{Cost Function}

The cost function is designed to avoid time-consuming long moves of the sample and to forbid re-measuring previously explored points. The practical parameter change proceeds by sequential moves in $x$, $y$ and $\Theta$. The first two correspond to a physical movement of the sample by piezoelectric motors, while the last is the adjustment of an electrostatic potential and thus so much faster than the physical movements that it does not have to be considered.  
The cost function is defined as
\begin{align}
f_c(\mathbf{x}, \mathbf{x}_{c}, \{\mathbf{p}_i\} ) = 
  \begin{cases}
      \infty,              & \text{if } \mathbf{x}\in \{\mathbf{p}_i\} \\ 
    1 + c_\mathrm{w} (\frac{d_{x}}{v_x} + \frac{d_{y}}{v_y})& \text{otherwise},
\end{cases}
  \label{eq:cost_function}
\end{align}
where  $v_{x}$, $v_{y}$ are the speeds to move the motor along the $x$ and $y$ directions, respectively, and $d_{x}$ and $d_{y}$ are the distances between the points calculated from $\mathbf{x}^{\prime}$ and $\mathbf{x}_{c}$. $c_\mathrm{w}$ is a parameter that can be used to adjust the relative weight of the movement time to the measurement time.

\section*{Results and Discussion}
\subsection*{Two-Dimensional Parameter Space}

The maps in Fig. \ref{fig:ground_truth}(a) and (b) show $\overline{I}$ and $\overline{C}$ measured by a conventional grid search on a grid with 5~$\mu$m spacing and $\approx$~22,700~measured points. Every point corresponds to a full $I(E,\Phi)$  spectrum. All spectra have been collected for a fixed $\Theta = 2^{\circ}$, which was chosen arbitrarily (the optimum choice of $\Theta$ depends on the detailed sample mounting and sample shape and cannot be known before performing the experiment). A few representative spectra are shown in Fig. \ref{fig:ground_truth}(c)-(f). The locations at which these have been taken are marked in both panel (a) and the panels of the spectra. This data set can serve as a ground truth for testing autonomous experiments in the two-dimensional $(x,y)$ parameter space.

 Before presenting these autonomous experiments, we briefly discuss the properties of the data set. One first notices a strong correlation between $\overline{I}$ and $\overline{C}$ displayed in Fig. \ref{fig:ground_truth}(a) and (b), respectively.  This has several reasons. First,  a high value of $\overline{C}$ necessitates a high $\overline{I}$ in the sense a minimal signal cannot show a high $\overline{C}$ (unless it is created by noise). Moreover, the maps in Fig. \ref{fig:ground_truth}(a) and (b) span a very large range of values, from no signal at all to high values, and more subtle changes are thus less pronounced on the colour scale. Nevertheless,  the entire upper right quarter of the images displays more structural differences in $\overline{C}$. Given the similarity between $\overline{I}$ and $\overline{C}$ in these maps, we will present the results from the autonomous experiments solely in terms of $\overline{C}$. However, including both tasks in the GPR has the added benefit of decreasing the sensitivity towards fluctuations in either task and adds stability to the autonomous experiment. 
 
 The detailed structure of the maps in Fig. \ref{fig:ground_truth}(a) and (b) depends on the choice of the ROI. Here this covers almost the entire angular range but the energies are restricted to a range from just below the valence band maximum to the Fermi level (see dashed rectangle in Fig. \ref{fig:ground_truth}(c)). The ROI thus mainly includes the topological surface state of Bi$_2$Se$_3$ (when present), the bottom of the conduction band and the top of the valence band \cite{Xia:2009aa,Bianchi:2010ab,Bianchi:2012ac}. The most pronounced features are a conical dispersion (a so-called Dirac cone) with a ``filled'' upper part, stemming from the filled conduction band minimum, typical for the degenerate $n$ doping of this material.  A systematic exploration of the topological surface state at a specific location is shown in the lower part of Fig. \ref{fig:setup_sketch} as a series of $I(E,\Phi)$ spectra for different angles $\Theta$. If the topological surface state is present, as in Fig. \ref{fig:ground_truth}(c) and (d), this causes both a high $\overline{I}$ and a high $\overline{C}$ but if it is not, the area typically shows a very low $\overline{I}$ and $\overline{C}$; see Fig. \ref{fig:ground_truth}(e), (f).

We observe strong differences in the individual $I(E,\Phi)$ spectra over the sample area. Fig. \ref{fig:ground_truth}(c) shows an almost ideal topological surface state with the upper and lower parts of the conical dispersion meeting in a single (Dirac) point marked by an arrow. In Fig. \ref{fig:ground_truth}(d), the spectrum looks very similar but a small gap is opened between the upper and the lower part of the cone and the Dirac point is not visible. The most likely reason is that the sample normal direction at this location is slightly misaligned with respect to the electron analyser; this could be compensated by changing the measured emission angle $\Theta$ perpendicular to $\Phi$. Indeed, the apparent opening of a gap near the Dirac point can also be observed in the spectra shown in Fig. \ref{fig:setup_sketch}. The Dirac point crossing is only observed for a perfect alignment and lost upon a very small change of $\Theta$. The spectrum in Fig. \ref{fig:ground_truth}(e) does not show the typical Dirac cone structure of Bi$_2$Se$_3$ but the strong feature at an energy of $\approx$1.2~eV appears quite similar to those in panels (c) and (d). As we shall see later, the reason for this is that the misalignment in this sample region is so large that the Dirac cone falls altogether out of the observed angular range. Finally, Fig. \ref{fig:ground_truth}(f) is taken from a point outside the sample area. There is still some intensity observed, presumably from stray light hitting the actual sample. 

From these observations it already becomes clear that the present two-dimensional scan is insufficient for a complete characterisation of the sample. For example, the area around the spectra of both Fig. \ref{fig:ground_truth}(e) and (f) shows low $\overline{I}$ and $\overline{C}$ but for the latter the reason is that we are outside the sample area while for the former it is that the sample is misaligned and we are thus merely probing the wrong $\Theta$ (as we shall see below). 

\begin{figure}[htbp]
  \includegraphics[width=\figwidth]{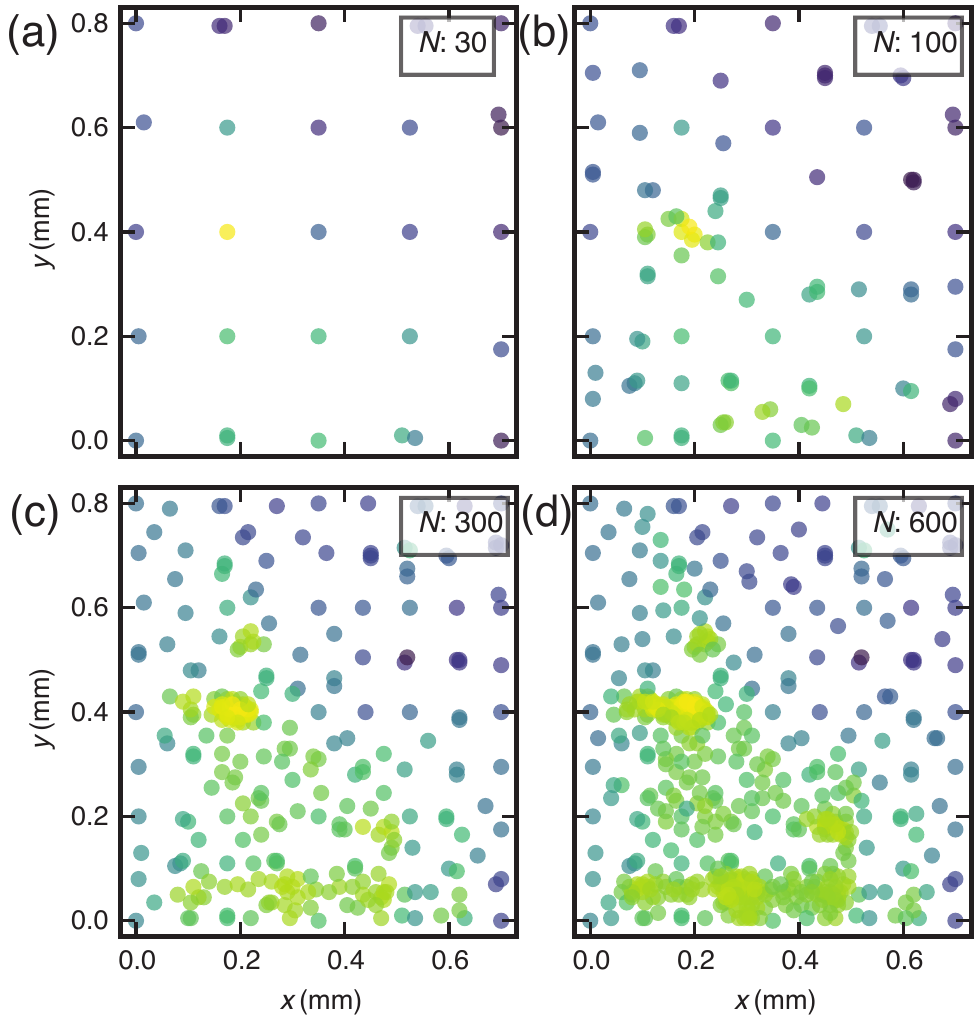}
  \caption{GPR-driven exploration. The points show the locations on the sample chosen by the GPR at different stages of the experiment. Panels (a) to (d) show the first $N=$ 30, 100, 300 and 600 data points, respectively. The colour of the points represents the mean curvature $\overline{C}$.
  }
  \label{fig:clustering_2D}
\end{figure}

The autonomous data acquisition was performed in the same area as in Fig. \ref{fig:ground_truth}(a). $\alpha$ in equation (\ref{eq:our_acqisition_function}) was chosen to favour exploitation rather than exploration, making the GPR-guided search distinctly different from random sampling. The result is illustrated in Fig. \ref{fig:clustering_2D} by plotting the initialisation points (18 points) together with the points chosen by the GPR after a total of $N$ collected points from the autonomous acquisition and re-training of the GP as outlined in the description of the GPR script. While both tasks ($\overline{I}$ and $\overline{C}$) where used for the autonomous search, the colour of the points is chosen to represent just $\overline{C}$ for simplicity.  At the beginning of the data acquisition, the points are typically set close to the boundaries and on a regular grid across the sample (Fig. \ref{fig:clustering_2D}(a)). Once more points are added, one starts to see that the GPR search targets areas of high acquisition function, leading to a higher point density there (panels (b) to (d)). From the colour of the data points, on can see that a relatively good representation of the overall sample landscape is already achieved for $N=300$ in panel (c) and a very good representation is found for $N=600$ in panel (d). For 300 and 600 data points, we observe a strong clustering of the points on the lower edge of the sample region and for ($x\approx$0.2~mm, $y\approx$0.4~mm). In these areas, the topological surface state is especially intense and sharp, as seen in Fig. \ref{fig:ground_truth}(c) and (d). 

\begin{figure}[htbp]
  \includegraphics[width=\figwidth]{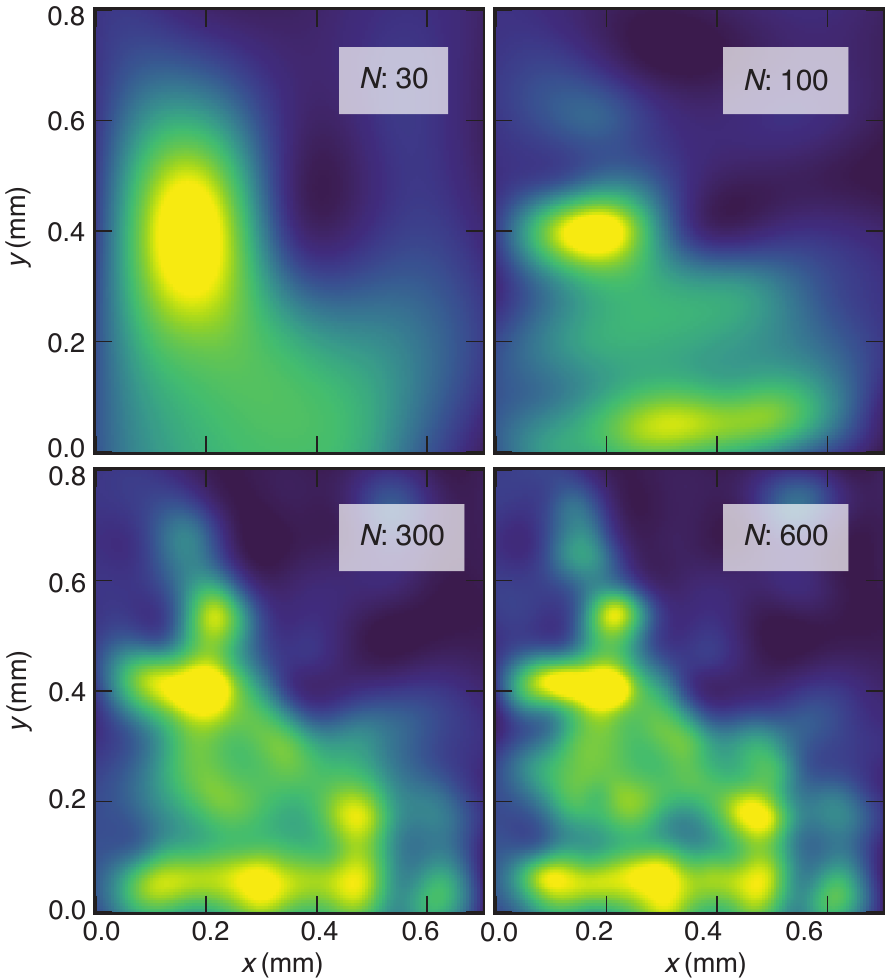}
  \caption{(a)-(d) Posterior distributions of $\overline{C}$, evaluated after collecting the first $N=$ 30, 100, 300 and 600 data points, respectively, and re-training the GP as described in the text.}
  \label{fig:posterior}
\end{figure}

As a test of the GPR-driven data acquisition, we can plot the landscape of the posterior task values across the sample surface and compare it to the ground truth of Fig. \ref{fig:ground_truth}. In Fig. \ref{fig:posterior}, this is plotted for the same number of data points as in Fig. \ref{fig:clustering_2D}. The posterior of the GP already shows the large-scale structure of the ground truth for $N=100$. At  $N=300$, it has a considerable amount of fine structure. While this continues to improve at $N=600$, the changes become small.  In any case, this result illustrates the capability of the approach to explore the sample with a very small number of measurements (compared to the 22,700 data points used to construct Fig. \ref{fig:ground_truth}).

\subsection*{Three-Dimensional parameter Space}

A search of the sample surface for specific electronic features using a hemispherical analyser can only be successful if these features are consistently appearing within the range of emission angles $\Phi$ covered by the two-dimensional detector of the electron analyser and are thus certain to be found in every measured spectrum $I(E,\Phi)$. In most practically relevant cases, this is not so: For any sample with an anisotropic electronic structure (\emph{i.e.}, not only free electron-like states), the azimuthal direction of the $\Phi$ angle with respect to the sample orientation is in principle arbitrary and there is no guarantee one will find features of interest at the Brillouin boundary, such as the Dirac point of graphene, in the detected $\Phi$ range. Even structures around the Brillouin zone centre (normal emission) might not be inside the detection range if the sample is tilted with respect to the sample-analyser direction. As we shall see, this is the case for parts of our sample. The spectrum in Fig. \ref{fig:ground_truth}(e), for instance, does not show the characteristic topological Dirac cone of Bi$_2$Se$_3$ but the only reason for this is that this particular crystalline domain is tilted so much as to move the Dirac cone completely out of the detection window. 

These issues can be solved by including three parameters in the scan, $(x,y)$ and the analyser angle $\Theta$ to cover a wide range of emission angles perpendicular to the $\Phi$ direction. In this way, the complete electronic structure in the two-dimensional momentum space is explored for each position of the sample. In a sense, such a three parameter scan is the ``minimal'' requirement for a comprehensive sample characterisation using spatially resolved ARPES and yet, to the best of our knowledge, it is not implemented at any of the operational micro- or nanoARPES beamlines. It is easy to see the importance of this approach: The map of the sample area in Fig. \ref{fig:ground_truth}(a) contains $\approx$22,700 points. Measuring this map for a range of $\Theta$ of $\approx$20\degree~in steps of 1\degree~in order not to miss any features would call for measuring a total of $\approx$477,000 points. Even with a dwell time as short as 0.5~s and the (unrealistic) assumption of no movement or overhead time between points, such a measurement would take several days -- something that is completely prohibitive for an exploratory search in a realistic beamtime schedule. 

\begin{figure}[htbp]
  \includegraphics[width=\figwidth]{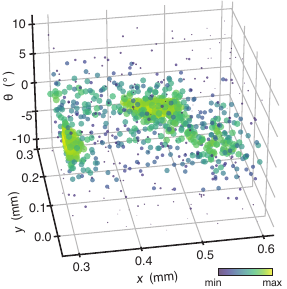}
  \caption{Results of a GPR-driven experiment ($N=$940) for three parameters: the sample position $(x,y)$ and the emission angle $\Theta$. Distribution of measured points. The marker size and the colour both encode the mean curvature $\overline{C}$. The points are partially transparent to improve the visualisation. }
  \label{fig:clustering_3D}
\end{figure}

In the following, we demonstrate the extension of the GPR to a third parameter dimension. We have limited this type of search to a smaller area on the sample surface, indicated by the red dashed rectangle in Fig. \ref{fig:ground_truth}(a). The emission angle $\Theta$ was allowed to vary between +10\degree~and -10\degree. The results of a search with $N=$940 unique positions is  given in Fig. \ref{fig:clustering_3D} as a three dimensional visualisation of the measured points. The point size and colour are both encoding the value of $\overline{C}$. Again, the results stem from a two-task GPR and the corresponding results for $\overline{I}$ look extremely similar.
The plot shows a clear clustering in different regions of the parameter space. Three major clusters are found on the left, centre and right hand side of the $x$ scan range.

\begin{figure}[htbp]
  \includegraphics[width=\figwidth]{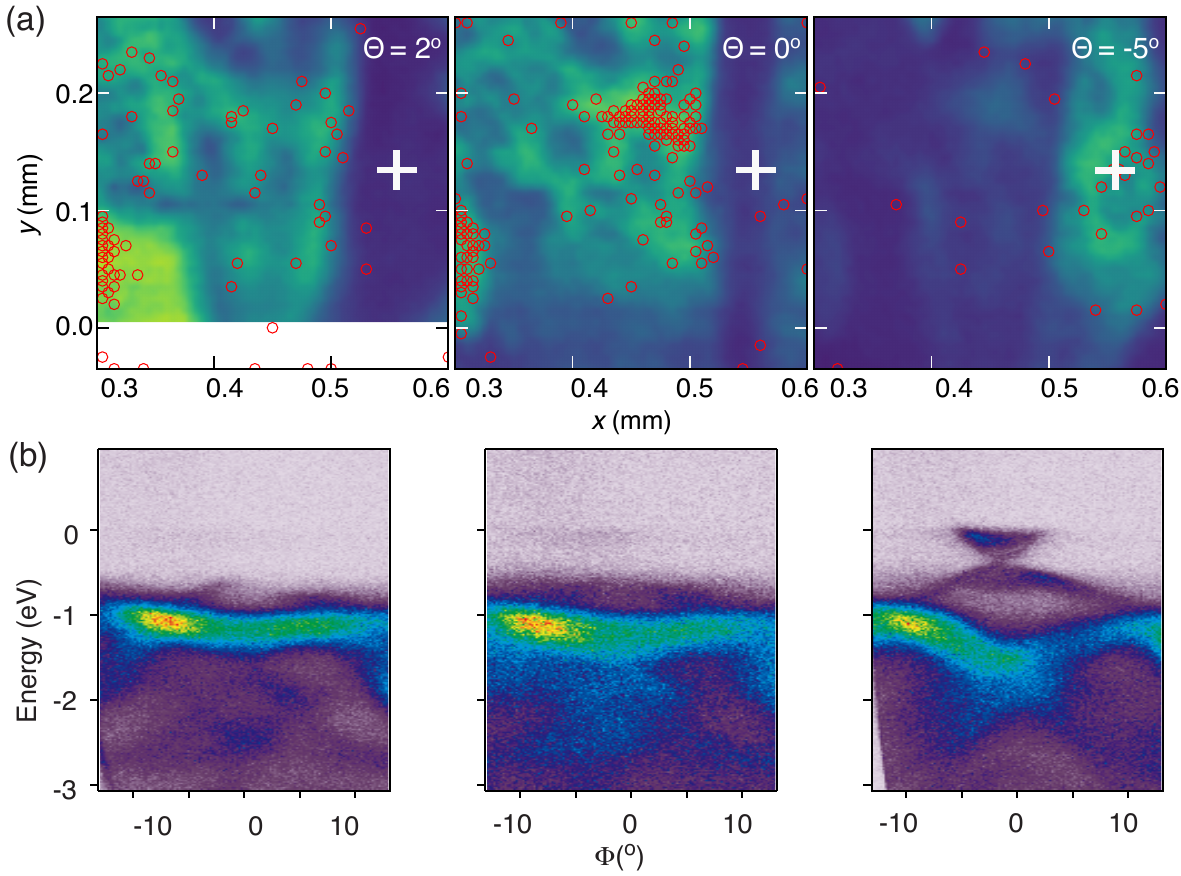}
  \caption{(a) Distribution of the data points from Fig. \ref{fig:clustering_3D} but only for angles $\Theta=2$, 0 and -5\degree. The background images are the mean curvature $\overline{C}$ from grid scans corresponding to that in Fig. \ref{fig:ground_truth}(b). The white cross markers indicate the position in which the spectra of panel (b) were taken. (b) $I(E,\Phi)$ spectra at the position marked in panel (a) (which is the same position as for Fig. \ref{fig:ground_truth}(e)) for the three values of $\Theta$ in the maps above. }
  \label{fig:dirac_cluster_3D}
\end{figure}

The clustering situation is further visualised in Fig. \ref{fig:dirac_cluster_3D}(a) by plotting the results only in the $\Theta=2$, 0 and -5\degree~planes. The measured positions are shown as red markers. The background images are ``ground truth'' grid scans for $\overline{C}$ at the three values of $\Theta$, taken with a grid spacing of 5~$\mu$m in the $(x,y)$ range of the three-dimensional autonomous experiment (the background for $\Theta=2$\degree~is a magnification of Fig. \ref{fig:ground_truth}(b)). The main cluster at $\Theta = 0$\degree~is found in the upper half of scan range. A second large cluster is more distributed in $\Theta$ and found at the left edge of the scan range in both the $\Theta=2$\degree~ and $\Theta=0$\degree~maps. The final cluster is found on the right edge of the scan range for $\Theta = -5$\degree. 

The number of identified clusters, invisible at fixed deflector angles, and the variety in the underlying maps and spectra of Fig. \ref{fig:dirac_cluster_3D}, illustrates the necessity of a three-dimensional autonomous parameter search, including a variation of $\Theta$. The data indicate that the sample consists of small Bi$_2$Se$_3$  domains of high crystalline quality, with slightly different orientations. The Dirac cone, which is the identifying electronic structure feature of the topological phase, has only a small width in angle at this photon energy (18~eV), see also Fig. \ref{fig:setup_sketch}. If the Dirac cone is well-centred in the $\Theta =-5$\degree~image, it is certain to be missed in a $\Theta =2$\degree~image, and vice versa. Indeed, the well-established Dirac cone in Fig. \ref{fig:dirac_cluster_3D}(b) for $\Theta =-5$\degree~is not observed in the spectra taken  for other values of $\Theta$ but at the same location. The position of the cross marker in Fig. \ref{fig:dirac_cluster_3D}(a) shows a corridor of low $\overline{I}$ and $\overline{C}$ in Fig. \ref{fig:ground_truth}(b) and, equivalently, in Fig. \ref{fig:dirac_cluster_3D}(a) for $\Theta = 0$ and 2\degree. Based only on a two-dimensional $\Theta=2$\degree~scan, the topological state at this location would be missed and the region might be mistakenly assigned to a different material composition. 

Finally, we note that the GPR-controlled experiment is well-tuned to identify the desired features in the sample. The regions where measurement point clustering occurs show the Bi$_2$Se$_3$ Dirac cone and the conduction band minimum, as illustrated by the spectra in Fig. \ref{fig:ground_truth}(c) and (d) that lie in the regions of strong clustering for the  two-dimensional autonomous experiment (see Fig. \ref{fig:clustering_2D}). Correspondingly, the spectrum for $\Theta=-5$\degree~in Fig. \ref{fig:dirac_cluster_3D}(b) appears in the smaller cluster at $\Theta=-5$\degree~in Figs. \ref{fig:clustering_3D} and \ref{fig:dirac_cluster_3D}(a). We stress again that clustering on Dirac-cone regions is closely linked to the definition of the ROI around the area of the Dirac cone in Bi$_2$Se$_3$ (see Fig. \ref{fig:ground_truth}(c)), illustrating the need to make a physically motivated choice in the process that, in this case, is a preference for the metallic parts of the sample. Note that the ROI is selected around the entire Dirac cone and we thus do not observe a distinction between a cone with a detectable Dirac point (Fig. \ref{fig:ground_truth}(c)) and a gap at the Dirac point (Fig. \ref{fig:ground_truth}(d)). Such a distinction could be made by a search with a very small ROI in energy centred around the Dirac point. 

\section*{Conclusion}

In conclusion, we have implemented autonomous GPR-controlled microARPES experiments at the SGM4 beamline of ASTRID2 and demonstrated this by exploring two- and three-dimensional parameter spaces. In particular the latter application is shown to be crucial for microARPES  because it implements the minimal search of the complete angular space for every position on the sample. The restriction to a two-dimensional spatial $(x,y)$ parameter space at a fixed angle would have caused missing important electronic structure features for our particular sample. The GPR-controlled search is also significantly increasing the efficiency of the experiment, requiring only a tiny fraction (a few percent at most) of the data to give a reasonable estimate of the task distribution in the investigated parameter space. This is especially important for high dimensional parameter spaces. Moreover, the case of a three-dimensional parameter space illustrates another advantage of the autonomous experiment: For a trained human expert, it is still relatively easy to explore four-dimensional data such as $I(E,\Phi)$ images on a two-dimensional $(x,y)$ grid. Given the right visualisation tools, looking through the $\approx$22,700 images generating Fig. \ref{fig:ground_truth} and identifying the most promising areas takes only a few minutes. For a higher dimensional parameter space, such as one that integrates an exploration of $\Theta$, this is increasingly difficult.

There are several possible improvements that can be easily implemented and tested using the approach to autonomous microARPES experiments introduced here. The GPR could be refined by other kernels or acquisition functions, or it could be substituted for an altogether different search strategy such as  reinforcement learning. A particularly interesting challenge is the implementation of new tasks that are more tuned towards physically relevant features across a large ROI. All developments need to retain the flexibility required to adapt to the many situations relevant in a spatially resolved ARPES experiment, in particular the great possible variation within sample makeup, quality and orientation. Problems can range from the search for the highest quality spot on a nearly uniform sample, over needle-in-the-haystack tasks of finding a particularly interesting but small region in a very large sample, to identifying the number and character of separated phases in a quantum material. Depending on the physical question, AI-based search strategies need to be adapted and combined with human expertise for an optimal outcome. 

\section*{Acknowledgments}
\begin{acknowledgments}
We thank Simone Kevy and Martin Bremholm for the Sb-doped Bi$_2$Se$_3$ crystals used in this study. This work was supported by VILLUM FONDEN via the Synergy programme (grant number 40558) and by the Danish Council for Independent Research, Natural Sciences under the Sapere Aude program (Grant No. DFF- 9064-00057B), and Grant No. DFF-1026-00089B.
  
   \end{acknowledgments}

\end{document}